\RequirePackage{fix-cm}
\documentclass{svjour3}                     
\smartqed  

\usepackage{graphicx}
\usepackage[numbers]{natbib}
\usepackage{siunitx}
\usepackage{dcolumn}
\usepackage{booktabs}
\usepackage{comment}
\usepackage{color}
\usepackage[normalem]{ulem}

\begin{document}

\title{Superconducting Niobium Calorimeter for Studies of Adsorbed Helium Monolayers
}


\author{Jun Usami$^1$         \and
       Koki Tokeshi$^1$ \and Tomohiro Matsui$^1$ \and Hiroshi Fukuyama$^{1,2}$ 
}


\institute{           J. Usami \at
              \email{jusami@crc.u-tokyo.ac.jp}           
          \and
           H. Fukuyama \at
              \email{hiroshi@kelvin.phys.s.u-tokyo.ac.jp
              \and
$^1$Department of Physics, The University of Tokyo, 7-3-1, Hongo, Bunkyo-ku, Tokyo 113-0033, Japan \\
          $^2$ Cryogenic Research Center, The University of Tokyo, 2-11-16, Yayoi, Bunkyo-ku, Tokyo 113-0032, Japan
}
}
\date{\today}

\maketitle

\begin{abstract}

We developed a calorimeter with a vacuum container made of superconducting niobium (Nb) to study monolayers of helium adsorbed on graphite which are prototypical two-dimensional quantum matters below 1\,K. 
Nb was chosen because of its small specific heat in the superconducting state.
It is crucially important to reduce the addendum heat capacity ($C_{\rm{ad}}$) when the specific surface area of substrate is small.
Here we show details of design, construction and results of $C_{\rm{ad}}$ measurements of the Nb calorimeter down to 40\,mK.
The measured $C_{\rm{ad}}$ was sufficiently small so that we can use it for heat capacity measurements on helium monolayers in a wide temperature range below 1\,K.
We found a relatively large excess heat capacity in $C_{\rm{ad}}$, which was successfully attributed to atomic tunneling of hydrogen (H) and deuterium (D) between trap centers near oxygen or nitrogen impurities in Nb.
The tunnel frequencies of H and D deduced by fitting the data to the tunneling model are consistent with the previous experiments on Nb doped with H or D.


\end{abstract}

\section{Introduction}
\label{intro}

Atomically thin $^4$He and $^3$He films physisorbed on flat surfaces at low temperatures provide experimental opportunities for us to study bosonic and fermionic quantum phenomena, respectively, in two-dimensions (2D).
Among them, a few layers of $^4$He and $^3$He adsorbed on graphite are of particular importance because of their rich quantum phase diagrams~\cite{Bretz1973,Greywall1990,Greywall1993,Godfrin1995,Fukuyama2008}.
Heat capacity measurements are one of the most sensitive experimental techniques to explore the phase diagrams in these systems.
One important unsolved problem in this field is the possible self-binding of $^3$He, i.e., 2D liquefaction of $^3$He, which was claimed by Sato $\it{et~al}$~\cite{Sato2010,Sato2012}.
They found that the uniform fluid state of areal densities lower than 0.6-0.9\,nm$^{-2}$ is unstable against phase separation into gas and liquid phases below 80\,mK, the gas-liquid (G-L) coexistence.
This theoretically unexpected finding stimulated new theoretical studies~\cite{Ruggeri2013,Ruggeri2016,Gordillo2016c,Gordillo2016b}, and some of them~\cite{Ruggeri2016,Gordillo2016b} can partially explain the experimental results of Sato $\it{et~al}$. but not sufficiently.
Similar experimental indications of the self-binding are also reported in 2D $^3$He on different substrates from graphite~\cite{Bhattacharyya1985,Csathy2002}.

In general, for studies of adsorbed 2D systems, comparison of heat capacity data obtained on substates with different surface coherence lengths ($\xi$) is crucially important for determining the true nature of detected phase transitions~\cite{Bretz1977}.
In the case of exfoliated graphite, the comparison can be done using Grafoil and ZYX substrates, where ZYX has a ten times longer $\xi$ ($\approx200$\,nm) than Grafoil which is used in the most previous studies including the measurements by Sato $\it{et~al}$.
Thus, a new heat capacity measurement with ZYX is highly desirable to test if their results vary quantitatively or even qualitatively due to the finite size effects.
In addition, it is important to observe the G-L critical point which may exist at temperatures $\it{above}$ 80\,mK, the high temperature limit of the previous surveys~\cite{Sato2010,Sato2012}.
Besides the 2D liquefaction of $^3$He, there are many other interesting physics to be explored by heat capacity measurements in 2D He systems~\cite{Morishita2001,Casey2003,Nakamura2016,Nyeki2017a}.

The heat capacity contribution from the empty calorimeter (addendum) with ZYX should be minimized to keep reasonably good experimental accuracies, because the specific surface area of ZYX ($= 2$\,m$^2$/g) is ten times smaller than that of Grafoil.
Previously, Nakamura $\it{et~al}$. constructed a calorimeter with a ZYX substrate in a container (sample cell) made of Nylon~\cite{Nakamura2012}.
Nylon is an easily machinable polymer with a sufficiently small specific heat below 1\,K~\cite{Pobell2007a}.
It worked well during the first several thermal cycles, but then started to leak because of many microcracks created by water absorbed inside the cell walls due to the high hygroscopy of Nylon.
In order to overcome this technical problem, we have constructed a new calorimeter with a vacuum container made of niobium (Nb), containing the same ZYX substrate as that used in the previous Nylon calorimeter.
Nb is a superconductor with the superconducting transition temperature ($T_{\rm{sc}}$) of 9.25\,K.
The specific heat of superconductors is known to fall off exponentially with decreasing temperature below $T_{\rm{sc}}$ and in proportion to $T^3$ below about $0.1T_{\rm{sc}}$ due to decreasing numbers of normal state electrons and thermal phonons, respectively.
Thus the addendum of Nb calorimeter is expected to be very small below 1\,K.

In this article, we report design details and measurement results of the addendum heat capacity of the Nb calorimeter.
It turned out that the measured addendum is as low as that of the previous Nylon calorimeter but is at least twice as large as the estimation based on the amounts of all construction materials of the cell and their known specific heats.
By comparing with a theoretical estimation for heat capacities due to atomic tunneling of hydrogen and deuterium impurities trapped by oxygen or nitrogen impurities in Nb, we concluded that the measured excess heat capacity comes from these contributions.

\section{Experimental}
\label{method}

There are three basic concepts in our calorimeter design. 
The first is to minimize the addendum heat capacity, of course.
The second is to increase thermal conductances among essential parts, such as the ZYX substrate, a thermometer and a heater, as high as possible to realize a short internal thermal relaxation time over the whole measurement temperature range.
The third is to achieve a tunability of the thermal conductance between the calorimeter and the base temperature plate of a refrigerator so that the same calorimeter can be used in a wide temperature ($T$) range from 1\,K to even below 1\,mK depending on the experiment.

To meet these requirements, we carefully chose the other construction materials, besides superconducting Nb, as well as the thermal contact and isolation methods.
For example, no stainless steel or BeCu parts such as bolts and nuts to tighten demountable thermal contacts were used in order to reduce the addendum heat capacity.
Instead, a specially manufactured alloy, Si$_{0.15}$Ag$_{0.85}$ (Tokuriki Honten Co., Ltd.), was used for this purpose. 
Most of the thermal links are made of high purity silver (Ag) (Tokuriki Honten Co., Ltd.; 99.999\%) rather than copper (Cu), since higher residual resistivity ratios (RRRs) of up to 3000 can more easily be obtained for Ag than Cu by a simple heat treatment at 630$^\circ$C for 2\,h in an oxygen pressure of 0.1\,Pa unless the size effect limits RRR.
The mechanism behind this heat treatment is the ``internal oxidation and segregation'' of magnetic impurities in noble metals~\cite{Ehrlich1974}.
Due to the higher RRR value, we could reduce the amount of Ag used.

To our knowledge, Nb has never been used as a main construction material in previously developed vacuum chambers for low-$T$ experiments.
This is presumably because it does not seem easy to make a leak-proof seal for Nb  which cannot be soldered.
Note that Nb has a very small thermal contraction~\cite{Wang1998} compared to typical epoxy sealants for cryogenic applications such as Stycast 1266 or 2850FT.
It was also unknown for us if commercially available Nb materials are helium leak-proof or not. 
Therefore, we first confirmed the leak-proofness of a prototypical Nb cell sealed with Stycast 1266 at $T = 77$\,K, and then constructed the actual sample cell.

\begin{figure}[t]
\centering
\includegraphics[width=\linewidth]{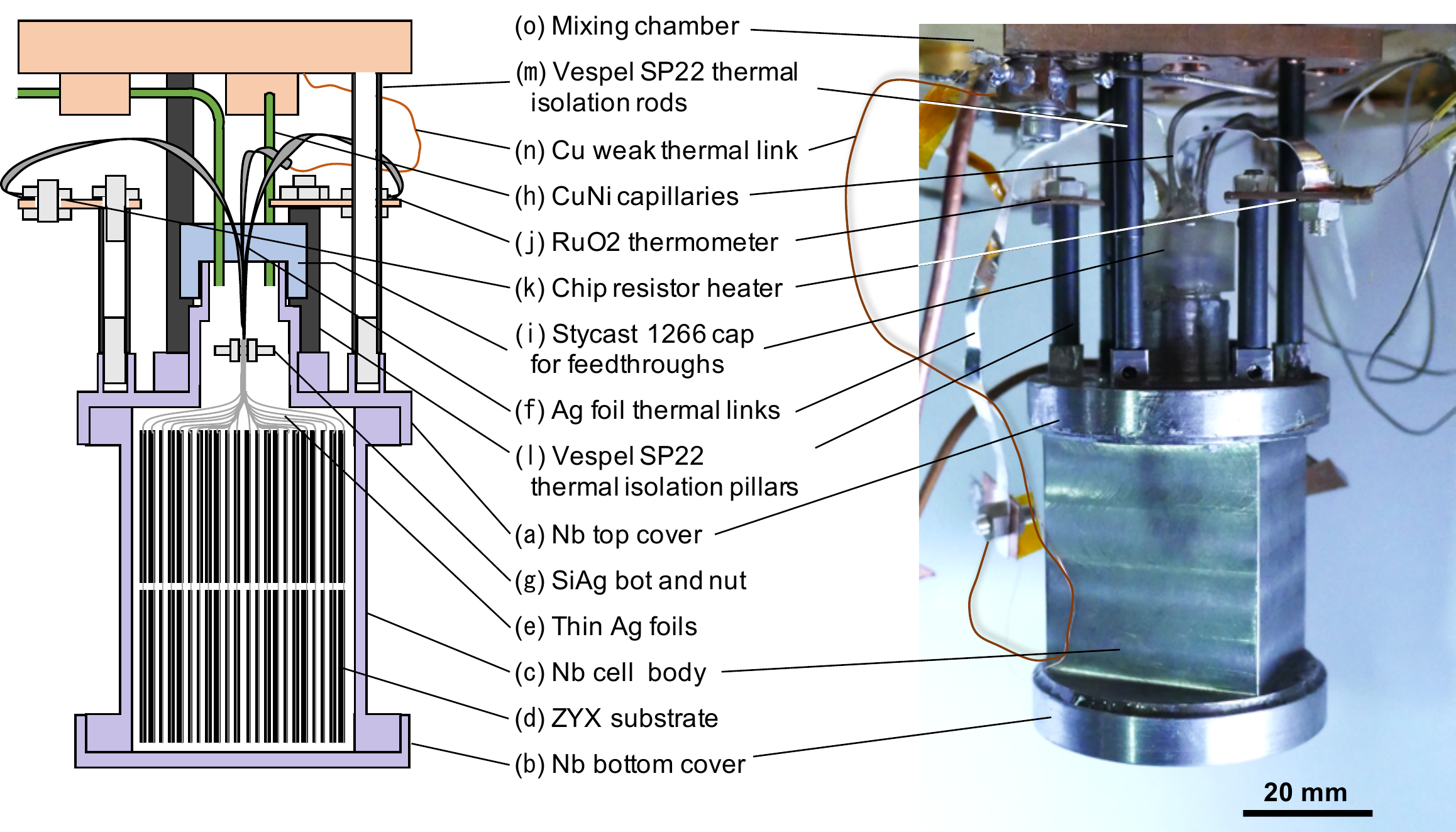}
\caption{\label{fig-cell}Schematic drawing ($\it{left}$) and photograph ($\it{right}$) of the Nb calorimeter  constructed in this work.}
\end{figure}

Fig.~\ref{fig-cell} shows a schematic drawing and a photograph of the calorimeter constructed in this work. 
The sample cell consists of three Nb parts: the (a)circular top and (b)bottom covers of 45\,mm in outer diameter and the (c)main cell body with a rectangular inner cross section of 22$\times$33\,mm$^2$.
They are machined from a single Nb rod (Changsha South Tantalum Niobium Co., Ltd.).
Typical impurity concentrations in the Nb rod ($> 99.93$at.\% purity, RRR $= 23$) provided by the supplier are C (37ppm), N (8ppm), O (95ppm), Si (1ppm), Fe (6ppm), Ni (7ppm), Mo (10ppm), Ta (460ppm), and W (12ppm).

The (d)adsorption substrate consists of ZYX plates (1\,mm thick each).
Every two such ZYX plates are diffusively bonded on both sides of a 50\,$\si{\micro}$m thick Ag foil to assist thermalization of the substrate. 
Details of the original substrate preparation are reported elsewhere~\cite{Nakamura2012}.
The ZYX substrate was taken out of the previous Nylon cell and was baked at 110$^\circ$C for 10\,h in vacuum.
Then, it was enclosed in the new Nb cell and sealed with Stycast 1266 in a glove box with a nitrogen (N$_2$) atmosphere.
The (e)thin Ag foils are tightly connected to (f)three thick Ag foils (0.2\,mm thickness) with the (g)SiAg bolt and nuts inside the cell.
The thick Ag foils and (h)two CuNi capillaries (0.6\,mm inner diameter) are fed through a (i)small Stycast 1266 cap on the top Nb cover which are sealed with Stycast 1266.
One of the CuNi capillaries is for sample filling, and the other is to connect the cell to a low-$T$ strain pressure gauge (not shown in the figure) to measure sample gas pressure $\it{in~situ}$.
After curing the bond, the calorimeter was taken out from the glove box and assembled underneath the mixing chamber plate of a dilution refrigerator as described below.
No He leak was detected from the Nb cell ($<3\times10^{-9}$\,ccSTP/s) when $^4$He gas of $1.3\times10^3$\,Pa was introduced to the cell at $T = 40$\,K.

Two of the three thick Ag foils thermally connect the (j)RuO$_2$ thermometer and the (k)heater to the ZYX substrate directly.
The last one connects the substrate with the mixing chamber through a weak thermal link.
The thermometer, which monitors the substrate temperature, is a RuO$_2$ chip register (Alps Electric Co., Ltd., 470\,$\rm{\Omega}$) wrapped in a silver foil of 50\,$\si{\micro}$m thickness.
This foil serves not only as a thermal link but also as a radiation shield for the register.
As the heater element for heat capacity measurements, we used a metal thin film resistor (Susumu Co., Ltd., 100\,$\rm{\Omega}$).
The thermometer and the heater are supported from the Nb top cover with (l)thermal isolation pillars made of Vespel SP22 (E. I. du Pont de Nemours and Company).

The Nb cell is supported from the mixing chamber plate with (m)three Vespel SP22 rods.
The (n)weak thermal link, which consists of two copper wires of 0.1\,mm in diameter and 10\,cm in length (RRR $ = 133$), connects between the ZYX substrate and the (o)mixing chamber with a moderate thermal conductance so that we can employ the quasi-adiabatic heat pulse method~\cite{Nakamura2013s} for heat capacity measurements in the temperature range of 40\,mK $\leq T \leq 1$\,K.
When the cell is empty, the measured internal thermal relaxation time is a few seconds, and the relaxation time between the substrate and the mixing chamber plate is 20 to 90\,s in this temperature range.
An ambient heat leak to the cell was measured as 10\,nW below 200\,mK.
Under this heat leak, the lowest attainable temperature of the substrate is 40\,mK, while the base temperature of the mixing temperature is 18\,mK.
We didn't employ a superconducting heat switch nor a mechanical heat switch not only to simplify the cell design but also to avoid large heat generation when they open at the lowest temperature.
In Table~\ref{table-mat}, we list amounts of all materials used in the present calorimeter.

\begin{table}[t]
\centering
\caption{\label{table-mat}Construction materials and their weights, which are used in the present Nb calorimeter. The symbols correspond to those in Fig.~\ref{fig-add}(b).}
\begin{tabular}{ccc}
   \toprule\toprule
    symbol & material &  weight (g)   \\
   \midrule
     A & niobium & 200 \\
     B & silver & 20 \\
     C & copper & 4.3 \\
     D & ZYX exfoliated graphite & 19 \\
     E & Stycast 1266 & 2 \\
    \bottomrule\bottomrule
 \end{tabular}
\label{table-mat}
\end{table}

\section{Results and discussion}
\label{result}

\begin{figure}[t]
  \begin{center}
   \includegraphics[width=0.9\linewidth]{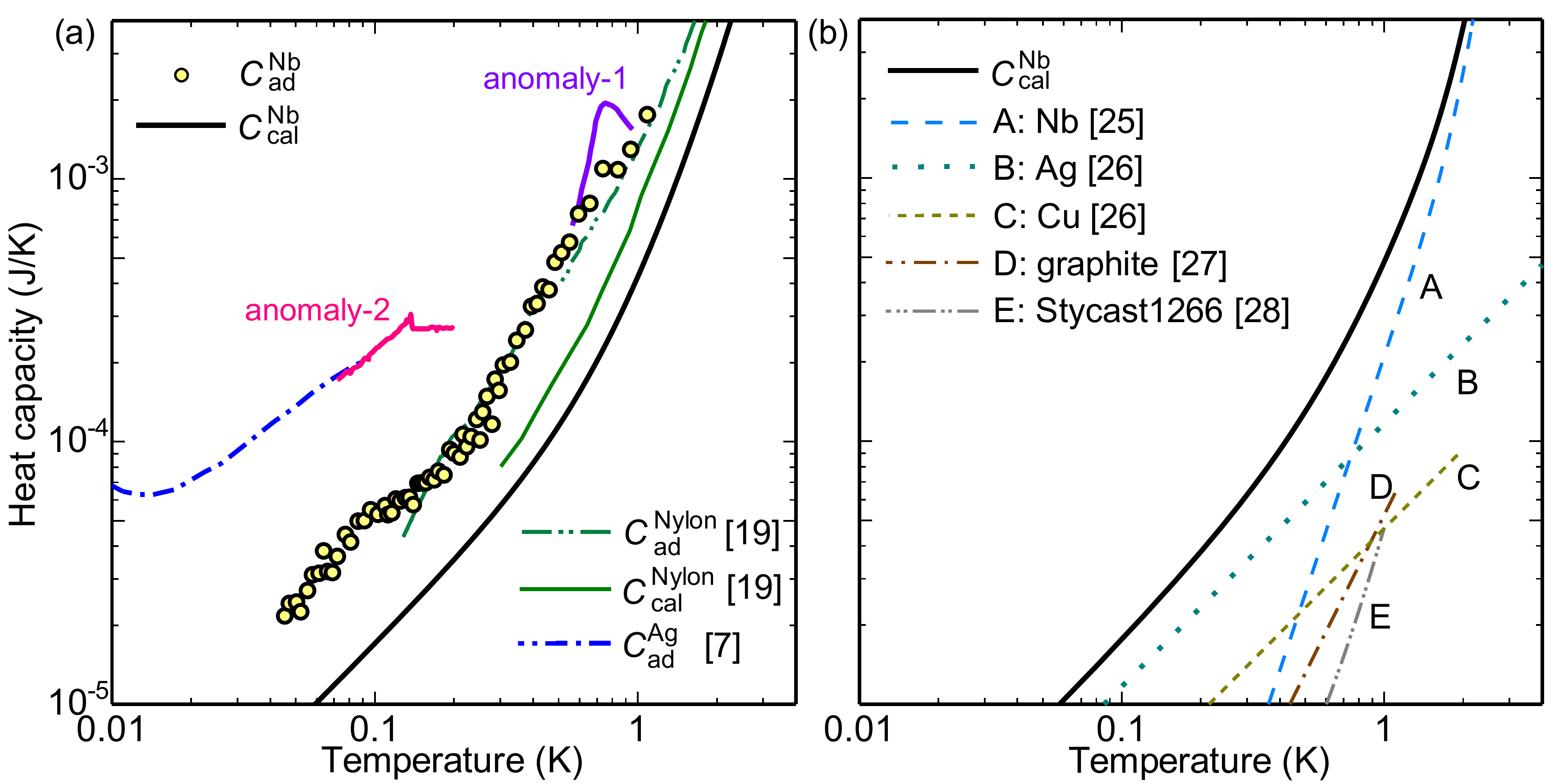}
  \caption{\label{fig-add}(a) Measured addendum heat capacities of the present Nb cell ((yellow) dots), the Ag cell ((blue) dashed dotted line~\cite{Sato2012}), and the Nylon cell ((green) dashed double-dotted line~\cite{Nakamura2012}) are compared. 
The heat capacity anomaly peaked at $T=0.7$\,K is associated with the G-L transition in the second layer of $^4$He on Grafoil~\cite{Greywall1993} (anomaly-1: (purple) line). The anomaly at $T=0.12$\,K observed in $^3$He submonolayer on a $^4$He superfluid thin film on Nuclepore is supposed to be related to the G-L transition~\cite{Bhattacharyya1985} (anomaly-2: (pink) line).  The (black) thick and (green) thin solid lines are addenda calculated from known properties of their construction materials in the Nb and Nylon cells, respectively. (b) Calculated heat capacity contributions from all construction materials shown in Table~\ref{table-mat} to the total one ((black) thick line) in the Nb cell. Here we used the published specific heat data for (A) Nb~\cite{Sellers1974}, (B) Ag~\cite{Martin1968}, (C) Cu~\cite{Martin1968}, (D) graphite~\cite{VanDerHoeven1966}, and (E) Stycast 1266~\cite{Siqueira1991}.}
    \end{center}
\end{figure}

Figure~\ref{fig-add}(a) shows the measured addendum heat capacity ($C_{\rm{ad}}$) of the present Nb cell ($C_{\rm{ad}}^{\rm{Nb}}$: (yellow) dots).
It has a steep temperature dependence at $T \ll T_c$ as is expected for superconductors.
Thus, $C_{\rm{ad}}^{\rm{Nb}}$ is much smaller than $C_{\rm{ad}}$ 
of the sample cell made of silver, a normal metal,  ($C_{\rm{ad}}^{\rm{Ag}}$: (blue) dashed dotted line~\cite{Sato2012}), where the specific heat
of conduction electrons falls slowly in proportion to the temperature at $T \leq 1$\,K.
In the figure, the peak drawn by the (purple) line at $T = 0.7$\,K (anomaly-1) is the heat capacity ($C$) anomaly associated with the G-L transition in the second layer of $^4$He adsorbed on Grafoil~\cite{Greywall1993} (areal density $\rho =$ 14 nm$^{-2}$) whose magnitude is normalized to be consistent with the surface area of our substrate.
Our Nb calorimeter would be able to detect this anomaly, since the peak height is larger than $C_{\rm{ad}}^{\rm{Nb}}$ by a factor of two at the same temperature.
The  kink at $T = 0.12$\,K drawn by the (pink) line (anomaly-2) in the figure is the $C$ anomaly associated possibly with the G-L transition in $^3$He submonolayer (0.092\,layers) floating on a 1.23\,nm-thick superfluid $^4$He film adsorbed on Nuclepore substrate~\cite{Bhattacharyya1985}.
$C_{\rm{ad}}^{\rm{Nb}}$ is about seven times smaller than this anomaly at the same temperature, which indicates that its detection would be much easier than the 0.7\,K anomaly.
As mentioned in Introduction, the possible G-L critical temperature ($T_{\rm{c}}$) in monolayers of $^3$He, which we are seeking for, is expected to emerge at $T \geq 80$\,mK.
The upper bound should be $T_{\rm{c}} = 740$\,mK in the second layer of $^4$He on Grafoil~\cite{Greywall1991}.
Note that $T_{\rm{c}}$ in $^3$He systems should be lower than in $^4$He systems due to larger zero-point energies.
Although the phase transition nature is unknown, the magnitude of the expected $C$ anomaly would not be so different from those of the two above-mentioned anomalies that were found previously in different 2D He systems.
Therefore, we expect to detect the possible finite-$T$ G-L transition of $^3$He in 2D with this Nb calorimeter which is immune from moisture brittleness unlike the Nylon cell.

As can be seen in Fig.~\ref{fig-add}(a), $C_{\rm{ad}}^{\rm{Nb}}$ is substantially larger than the calculated addendum heat capacity ($C_{\rm{cal}}^{\rm{Nb}}$; (black) thick solid line) in the whole temperature range we studied.
We denote the excess heat capacity as $C_{\rm{ex}}^{\rm{Nb}} = C_{\rm{ad}}^{\rm{Nb}} - C_{\rm{cal}}^{\rm{Nb}}$.
Here, $C_{\rm{cal}}^{\rm{Nb}}$ is evaluated from the known properties of all construction materials used in the cell (see Table~\ref{table-mat}). 
The contribution from the Nb cell walls to $C_{\rm{cal}}^{\rm{Nb}}$ is dominant at $T \geq 0.7$\,K, while that from the Ag parts is dominant at lower temperatures (see Fig.~\ref{fig-add}(b)).
The Nylon cell seems to have a similar but slightly smaller excess heat capacity ($= C_{\rm{ad}}^{\rm{Nylon}} - C_{\rm{cal}}^{\rm{Nylon}}$) than $C_{\rm{ex}}^{\rm{Nb}}$. 
In Fig.~\ref{fig-add}(a), the measured addendum heat capacity $C_{\rm{ad}}^{\rm{Nylon}}$ and the calculated one $C_{\rm{cal}}^{\rm{Nylon}}$ are also shown by the (green) dashed double-dotted line and the (green) thin line, respectively~\cite{Nakamura2012}.

\begin{figure}[t]
  \begin{center}
   \includegraphics[width=0.8\linewidth]{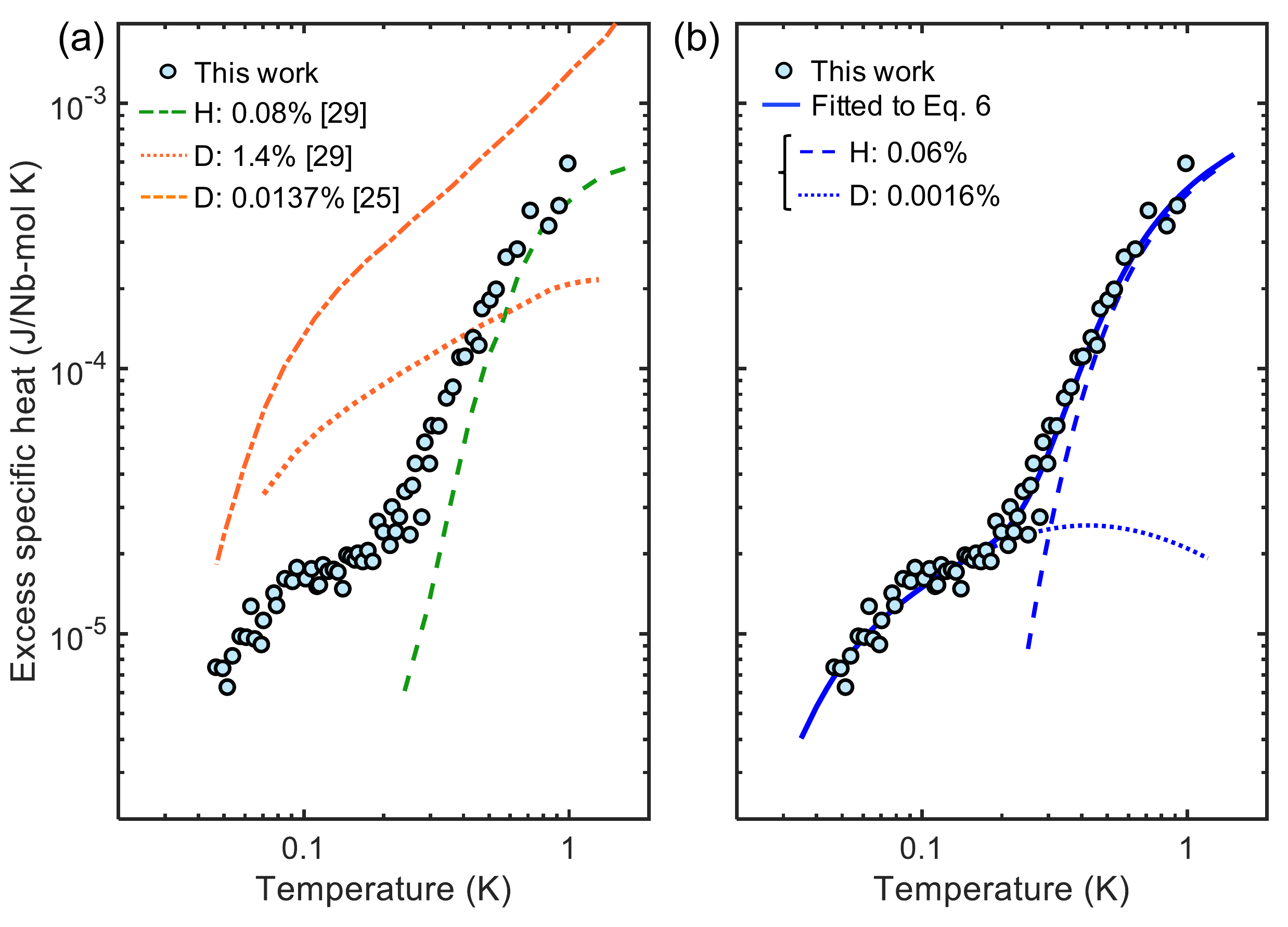}
  \caption{\label{fig-Cex}(a) Excess specific heat data ($c_{\rm{ex}}$: dots) of the present Nb cell are plotted with the $c_{\rm{ex}}$ data of Nb doped with hydrogen (H) at 0.08\% ((green) dashed line~\cite{Wipf1984}) and with deuterium (D) at 0.0137\% ((orange) dotted line~\cite{Sellers1974}) and 1.4\% ((orange) dashed dotted line~\cite{Wipf1984}) measured by the previous workers. (b) Fitting of the $c_{\rm{ex}}$ data (dots) to Eq.~\ref{eq-fit} (solid line). The fitting parameters are given in Table~\ref{table-fit}. The dashed and dotted lines are contributions from the H- and D-tunneling terms in Eq.~\ref{eq-fit}, respectively.}
    \end{center}
\end{figure}

The observed large $C_{\rm{ex}}^{\rm{Nb}}$ can be explained by atomic tunneling of hydrogen (H) or deuterium (D) impurities trapped by trap centers such as oxygen (O) or nitrogen (N) in Nb~\cite{Wipf1984}.
Figure~\ref{fig-Cex}(a) shows the excess specific heat of our Nb cell $c_{\rm{ex}}$, i.e., $C_{\rm{ex}}^{\rm{Nb}}$ divided by the molar amount of Nb used in the calorimeter, ((blue) dots).
The data seem to consist of two components.
One component, which dominates $c_{\rm{ex}}$ at high temperatures, rapidly falls down with decreasing.
The other one dominates below 0.2\,K.
Also shown in Fig.~\ref{fig-Cex}(a) are excess specific heats obtained by the previous workers for Nb samples with H of 0.08at\% ((green) dashed dotted line~\cite{Wipf1984}), D of 0.0137at\% ((orange) dashed line~\cite{Sellers1974}) and 1.4at\% ((orange) dotted line~\cite{Wipf1984}) introduced after high-vacuum annealing.
Here the concentrations of H and D in the data of Ref.~\cite{Wipf1984} are those determined by the authors themselves, and those of Ref.~\cite{Sellers1974} are obtained by fitting their data by ourselves with functional forms described below.
Note that the measured specific heat of the Nb sample after the high-vacuum annealing at 2250$^\circ$C in Ref.~\cite{Sellers1974} follows almost exactly $C \propto T^3$ expected from the Debye model (lattice vibrations).
All the data shown in Fig.~\ref{fig-Cex} are plotted after subtracting the $T^3$ contribution.
As can be seen in the figure, the high-$T$ component of our data is likely due to the H tunneling and the low-$T$ one due to the D tunneling.

Based on the above-mentioned scenario, we analyzed our data, $c_{\rm{ex}}(T)$, quantitatively following the method described in Ref.~\cite{Wipf1984} as follows.
The energy splitting $E$ of the ground state for a particle in an asymmetric double well potential is given by
\begin{equation}
E^2=\left(J^{2}+\epsilon^{2}\right), 
\label{eq-E}
\end{equation}
where $J$ is the tunnel frequency, and $\epsilon$ is the potential energy difference between the two minima caused by a random arrangement of neighboring O(N)-H or O(N)-D complexes.
Such $\epsilon$ has a distribution approximated by the Lorentzian distribution function $Z(\epsilon)$~\cite{Stoneham1969}:
\begin{equation}
Z(\epsilon)= \frac{\epsilon_{0}}{\pi (\epsilon_{0}^{2}+\epsilon^{2})}.
\label{eq-Z}
\end{equation}
Here the width $\epsilon_{0}$ is the typical energy difference of the double well. 
The specific heat of this tunneling system per one O(N)-H(D) complex, $c_{\mathrm{TS}}$, can be written as
\begin{equation}
c_{\mathrm{TS}} = k_{\rm{B}} \int_J^{\infty} d E Z(E) c(E,T),
\label{eq-cTS}
\end{equation}
where
\begin{equation}
Z(E)=\frac{2 E \epsilon_{0}} {\pi \left(E^{2}-J^{2}\right)^{1 / 2} \left(\epsilon_{0}^{2}+E^{2}-J^{2} \right)},
\label{eq-Z}
\end{equation}
\begin{equation}
c(E,T)=\left(\frac{E}{k_{\rm{B}} T}\right)^{2} \frac{\exp \left(-E / k_{\rm{B}} T\right)}{\left[1+\exp \left(-E / k_{\rm{B}} T\right)\right]^{2}}
\label{eq-c}
\end{equation}
and $k_{\rm{B}}$ is the Boltzmann constant.
Finally, we fit our $c_{\mathrm{ex}}$ data to the following expression:
\begin{equation}
c_{\mathrm{TS}} =\sum_{i=\rm{H,D}} n_i N_{\rm{A}} k_{\rm{B}} \int_{J_i}^{\infty} d E Z_i(E) c_i(E,T),
\label{eq-fit}
\end{equation}
where $N_{\rm{A}}$ is Avogadro's number, and $n_{\rm{H}}$ and $n_{\rm{D}}$ are the concentrations of O-H and O-D complexes in Nb, respectively.
Here, $i$ stands for H or D.

\begin{table}[t]
\centering
\caption{\label{table-parameter}Physical parameters obtained by fitting the $c_{\rm{ex}}$ data of our Nb calorimeter to Eq.~\ref{eq-fit}. Also shown are parameters obtained from the $c_{\rm{ex}}$ data by other workers~\cite{Sellers1974,Wipf1984}. The parameters of Ref.~\cite{Sellers1974} are the results we obtained by fitting their $c_{\rm{ex}}$ data in the same manner as for our data, whereas those of Ref.~\cite{Wipf1984} are results of their own analyses where $n_{\rm{H}}$ and $n_{\rm{D}}$ are fixed at those estimated from the measured RRR values.}
 \begin{tabular}{r@{.}l  r@{.}l  r@{(}l  c  r@{.}l  r@{.}l  r@{}l  l}
  \toprule\toprule
      \multicolumn{2}{c}{$J_{\rm{H}}$ {[}K{]}}   &  \multicolumn{2}{c}{$n_{\rm{H}}$ {[}\%{]}}   &   \multicolumn{2}{c}{$\epsilon_{0,\rm{H}}$ {[}K{]}}   &&  \multicolumn{2}{c}{$J_{\rm{D}}$ {[}K{]}}  &  \multicolumn{2}{c}{$n_{\rm{D}}$ {[}\%{]}}   &  \multicolumn{2}{c}{$\epsilon_{0,\rm{D}}$ {[}K{]}}  &  Ref.    \\  \midrule
     1&85(9)  &  0&06(2)  & 12&5)    &&  0&11(9)    &      0&0016(5)   & 0&.9(5)  & This work \\
     \multicolumn{14}{c}{}\\
       1&5(1)  &  0&017(3)  & 5&1) &&  \multicolumn{2}{c}{---}  & \multicolumn{2}{c}{---} & \multicolumn{2}{c}{---} & \cite{Sellers1974} \\
       \multicolumn{2}{c}{---}  &  \multicolumn{2}{c}{---}  &  \multicolumn{2}{c}{---}     &&   0&2(1)  &  0&0137(4)  & 2&.9(3) & \cite{Sellers1974} \\
      \multicolumn{14}{c}{}\\
      1&8(6) & 0&08 &  21&4)  &&     \multicolumn{2}{c}{---}  &    \multicolumn{2}{c}{---}     &      \multicolumn{2}{c}{---}          & \cite{Wipf1984}\\
      2&0(2) & 0&24 &  58&2)  &&   \multicolumn{2}{c}{---}  &    \multicolumn{2}{c}{---}    &       \multicolumn{2}{c}{---}         & \cite{Wipf1984}\\
         \multicolumn{2}{c}{---}   &   \multicolumn{2}{c}{---}   &   \multicolumn{2}{c}{---}   &&  0&21(4) &  0&05 & 10&(10) &  \cite{Wipf1984}\\
         \multicolumn{2}{c}{---}   &    \multicolumn{2}{c}{---} &  \multicolumn{2}{c}{---}  &&  0&2(1) & 0&26 &  33&(1) &  \cite{Wipf1984}\\
         \multicolumn{2}{c}{---}   & \multicolumn{2}{c}{---} &  \multicolumn{2}{c}{---} &&  0&2(3) & 1&4 & 88&(7) &  \cite{Wipf1984}\\
    \bottomrule\bottomrule
 \end{tabular}
\label{table-fit}
\end{table}

The fitting result is shown as the (blue) solid line in Fig.~\ref{fig-Cex}(b), and the fitting parameters are summarized in Table~\ref{table-fit}.
In the whole temperature range, the fitting quality is satisfactory.
The deduced $J$ values, $J_{\rm{H}}=1.84\pm0.09$\,K and $J_{\rm{D}}=0.11\pm0.09$\,K, are in reasonable agreement with the previous workers' results: $J_{\rm{H}}=2.2\pm0.2$\,K and $J_{\rm{D}}=0.24\pm0.02$\,K obtained by heat capacity measurements by Wipf and Neumaier (WN)~\cite{Wipf1984} and neutron spectroscopy measurement~\cite{Wipf1981}.
The larger error bar in our $J_{\rm{D}}$ estimation is due to the much lower deuterium concentration $n_{\rm{D}}$ in our as-received Nb sample compared to those in the previous workers' samples into which the D impurities were intentionally doped after cleaning by the high-vacuum annealing.
It is known that $J_i$ is independent of $n_i$, whereas $\epsilon_{0}$ increases rapidly with increasing $n_i$~\cite{Wipf1984}.
Our data are consistent with these trends.
Overall, it is strongly suggested that the measured excess heat capacity in the addendum of our calorimeter is due to atomic tunneling of hydrogen isotopes embedded in the Nb parts.

\section{Conclusions}

We developed the calorimeter that is mainly made of Nb for the study of low temperature  properties of He monolayers adsorbed on a ZYX graphite substrate.
The use of Nb is because of its small contribution to the addendum heat capacity, $C_{\rm{ad}}$, in the superconducting state.
The measured moderate thermal relaxation time, lowest attainable temperature (40\,mK) and the small $C_{\rm{ad}}$ meet the requirements to detect subtle heat capacity anomalies associated with phase transitions in adsorbed He systems in the temperature range between 40\,mK and 1\,K.
One promising application is the possible finite-$T$ gas-liquid transitions in $^3$He monolayers expected to be observable at temperatures between 80 and 740\,mK.

We also found the additional heat capacity ($C_{\rm{ex}}$), an excess over the expected addendum heat capacity that is calculated from the known specific heats of construction materials.
$C_{\rm{ex}}$ can be explained by atomic tunneling of hydrogen (H) and deuterium (D) impurities in Nb.
The tunneling frequencies of H and D obtained by fitting the $C_{\rm{ex}}$ data to this model are consistent with those obtained in the previous studies.
If one can anneal Nb parts in a high vacuum at $T = 2200^\circ$C before assembling the cell, $C_{\rm{ad}}$ would be further reduced by a factor of three providing a further improvement in the resolution of heat capacity measurement.

\begin{acknowledgements}

\ 
We are grateful to Sachiko Nakamura for helpful discussions and sharing her technical experiences on the construction of the previous Nylon calorimeter with us. 
The authors appreciate Megumi A. Yoshitomi for her contributions to the early stage of this work. 
We also thank Ryo Toda and Satoshi Murakawa for their valuable discussions and the machine shop of the School of Science, the University of Tokyo for machining the Nb calorimeter.
This work was financially supported by JSPS KAKENHI Grant Number JP18H01170. 
J.U. was supported by Japan Society for the Promotion of Science (JSPS) through Program for Leading Graduate Schools (MERIT) and Grant-in-Aid for JSPS Fellows JP20J12304.

\end{acknowledgements}

\bibliographystyle{spphys}       
\bibliography{Ref.bib}   

%
%

\end{document}